## Title

Normative active inference: A numerical proof of principle for a computational and economic legal analytic approach to AI governance

## Authors

Axel Constant[1*]
Mahault Albarracin[2,3,4,5]
Karl J. Friston[2,3]

## Institutions

1. Department of Engineering and Informatics, University of Sussex, Brighton, UK
2. VERSES, Los Angeles, CA, USA
3. Laboratoire d'Aalyse Cognitive de l'Information Université du Québec à Montréal, Montréal, CAN
4. School of Computing and Digital Technologies, Sheffield Hallam University, Sheffield, UK
5. Institut de santé et société, Université du Québec à Montréal, Montréal, CAN
6. Queen Sq Institute of Neurology, University College London, London, UK

## Correspondence*

axel.constant.pruvost@gmail.com
University of Sussex, School of Engineering and Informatics, Chichester I, CI-128, Falmer, Brighton, BN1 9RH, United Kingdom

## Acknowledgement

AC was supported by a European Research Council, Synergy Grant (XSCAPE) ERC-2020-SyG 951631.

## Software specification note

The simulations can be reproduced using the demo DEM_Law and the SPM_MDP_VB_XXX function of the latest SPM12 MATLAB toolbox.

## ABSTRACT

This paper presents a computational account of how legal norms can influence the behavior of artificial intelligence (AI) agents, grounded in the active inference framework (AIF) that is informed by principles of economic legal analysis (ELA). The ensuing model aims to capture the complexity of human decision-making under legal constraints, offering a candidate mechanism for "agent governance" in AI systems, that is, the [auto]regulation of AI agents themselves rather than human actors in the AI industry. We propose that lawful and norm-sensitive AI behavior can be achieved through "regulation by design," where agents are endowed with intentional control systems, or behavioral "safety valves," that guide real-time decisions in accordance with normative expectations. To illustrate this, we simulate an autonomous driving scenario in which an AI agent must decide when to yield the right of way by balancing competing legal and pragmatic imperatives. The model formalizes how AIF can implement *context-dependent preferences* to resolve such conflicts, linking this mechanism to ELA's conception of law as a scaffold for rational decision-making under uncertainty. We conclude by discussing how context-dependent preferences could function as safety mechanisms for autonomous agents, enhancing lawful alignment and risk mitigation in AI governance.

# 1 INTRODUCTION

This paper provides a computational account of the influence of norms, and specifically legal norms, on artificial intelligence (AI) agents based on the theory of active inference (AIF) in the life sciences[1], such as applied to robotics[2,3], and implementing principles of economic legal analysis (ELA) (for a review, see[4,5]). Normativity is the capacity of an agent to act according to imperatives defining what it "should" do. In AIF, this can stem from the nature of the agent's algorithm thereby imposing a form of intrinsic or perceptual (cf.[6]) normativity, where the source of the "should" comes from within (i.e., the agent "should" adopt variational free energy minimizing perception and conduct[7,8]). Normativity in AIF can also stem from the structure of the model and its parameterization, which—given the nature of the algorithm—will lead to the selection of behaviors that overtly comply, or not, to extrinsically defined norms, such as social, cultural or legal norms. Our computational model—using standard AIF algorithms—presumes intrinsic normativity and is designed to illustrate the capacity of AIF to generate behavior conforming to "extrinsically" defined norms

We defend the notion that AIF furnishes a good candidate model for the governance of artificial intelligence (AI) agents. AI governance splits into two broad domains of governance: (i) "agent" governance, or the governance of AI agents software sold on the AI market, and (ii) "actor" governance, or the governance of actors in the AI industry, such as deployers and providers of AI software[9]. From the point of view of agent governance, the mitigation of AI risks should be achieved by endowing AI agents with the ability to make autonomous decisions that align with human conceptions of the lawful conduct. This, we believe, can be achieved by endowing the agent with the ability to align its behavior to norms of conduct such as legal norms, an approach akin to "regulation by design"[10,11]. In the context of agent governance, regulation by design involves ensuring that AI systems are endowed with "safety valve" behavioral management mechanisms, acting in real time to allow AI agents to act lawfully like we do, more often than not. The upshot of our approach is to structure AI agent's decision making in a way that resembles that of human decision making based on the law, in hope of enabling alignment with human intents[12].

To achieve this, we describe a behavioral AIF model implementing principles of ELA. Rational action—according to the view of ELA adopted in this paper—is about making a choice that maximizes preferences, given the various preference-shaping contexts imposed by the law. We illustrate our model with a simulation scenario in which an autonomous agent controlling a self-driving car has to decide when to "yield the right of way" by steering into the right lane. This decision has to be made by adjudicating between competing behaviors under the law: crossing over a full line to yield the right of way to an emergency vehicle—and risking being honked at—or keeping one's lane until the line becomes dashed.

We show how AIF solves this problem using the construct of "context dependent preferences". Section 2 of this paper introduces how this construct can be grounded in ELA, thereby providing validity to the construct. Section 3 of this paper presents a simulation study to support the face validity of the construct of context dependent preferences as applied to legal normative decision making in AI agents. We conclude in section 4 with a discussion of the potential of AIF and the construct of context dependent preferences for "safety valve" mechanisms in autonomous AI systems.

# 2 ELA AND CONTEXT DEPENDENT PREFERENCES IN AIF

## 2.1 ELA

ELA theory starts from the assumption that human agents are rational, which can be summarized by the idea that their rationale for action is the maximization/satisfaction of their preferred outcomes[13]. ELA assumes that the norms imposed by legal rules factor in human decision-making processes through various mechanisms influencing beliefs about the possibility and conditions of realization of preferred outcomes. One view is that legal rules function as good heuristics for decision making when certain constraints apply, and adhering to a rule may be a good enough strategy to improve decision making.

Decisions are costly, and the more complex a decision is, the greater the cost of deliberation. If the cost of deliberation outweighs the cost of following the rule — i.e., if it's cheaper to follow the rule — the agent will follow the rule. Resorting to legal rules in that way may be useful, for instance when having to make decisions for our welfare that involve the outcome of other agents' decisions (e.g., "if I pay my taxes and we all do, I will benefit from utility services, but if I do not want to pay for taxes, I might have to figure how to get running water myself, therefore the action of paying taxes might be the most economically rational").

EAL applies to different issues in legal theory[4], and centrally to our purpose, to the question of what makes actions influenced by legal rules economically rational (i.e., preference maximizing). EAL grounds its view of the way the law weighs into rational decisions on the theory of expected utility, which holds the general view that rational actors maximize preferred outcomes weighted by their probability; that is, by accounting for the uncertainty in the distribution of possible outcomes. This conception of rational choice can be "thick" or "thin" depending on the extent to which one defines the object of the preference; a thin theory simply asserts that behavior maximizes preferences, and a thick theory attributes specific content to those preferences, in order to make the theory testable[14].

In ELA, the term preference is meant in a technical sense, as a "linear order R over some domain D"[15], that is, as a ranking of properties in a certain domain. For instance, someone may have a preference ranking R over properties of cars, such as "air conditioning" over "paint color", which will factor into the decisions on the domain D of car purchase. Preferences are thus attributes of observable or cognizable properties that matter in a domain of choice, not of a decision outcome per se (e.g., of the car I bought). There are debates over the exact nature of the influence of the law on preferences[4,15,16]. One can think of at least three possible positions: (i) that the law is "preference" shaping, (ii) that the law is "belief" shaping, or (iii) that the law is "context" shaping.

**(i) Preference shaping**: One can argue that the law changes one's order of preferences (e.g., "I prefer the blacked-out windows option over the AC option, but because blacked out windows are illegal, I now prefer the AC option"). An issue with this view is that although it is true that, over development, one can acquire preferences that reflect the prescriptions of the law (e.g., starting to prefer alcoholic beverages once passed the age of majority: a.k.a. Acquired taste[17]), it is unintuitive to think that the law acts through shifting preferences in real time (e.g., a teenager starting to prefer alcoholic beverages over a chocolate bar upon learning that a new bill brings down drinking age from 18 to 14).

**(ii) Beliefs shaping**: One can argue that the law changes the level of credence over the relationship between beliefs and preferences, or over the "likelihood" of preferred outcomes (e.g., "I now prefer AC over blacked out windows because the probability that my car with the blacked-out window gets delivered on time is now very low due to the new tariff bill"). In this case, the law influences preference maximizing behavior by influencing beliefs about the success of preference maximizing action. Here, the legal rule shapes behavior by shaping beliefs about the likelihood of preferred outcomes under different actions, not the preference ranking per se for the outcomes of action. This is conceivable but lacks the flexibility to account for the coexistence of competing beliefs-outcome mappings that may apply, counterfactually, depending on context, and for which one might hold differentiated preferences (e.g., "I will prefer AC "unless" an agreement is reached between countries").

**(iii) Context shaping**: One can argue that the law changes the context under which different outcomes can be preferred[18], thereby changing the preference order driving the preference maximizing behavior. On this view, a legal rule — especially one that has a permissive form (vs. a prohibitive form) — may allow for the deliberation over different courses of actions based on the context established by the rule. For instance, a taxation mechanism over the emission of greenhouse gases, or a credit system in the matter, may set a threshold for the maximum greenhouse gases emission allowed to companies, the threshold functioning as context. Such a rule offers a context in which firms that prefer maximizing profit may do so up until they meet their emission threshold, as the penalty for exceeding the threshold may not outweigh the profit that comes with increasing production. In this case, the firm maintains the same preference rankings over profit margins, but adjusts which preference ranking will guide their action, depending on where they sit with respect to the context set by the law.

For an AI agent to respond to the law in the way humans do, this AI agent would have to be endowed with the ability to contextualize their behavior under the law in one of the three ways the law shapes decision making. The third option, context shaping, is interesting because, on the one hand, it allows for multiple sets of preferences to coexist, and to be enacted depending on different layers of legal context. On the other hand, it does not require an account of how the law changes preferences or beliefs. It simply requires an account of the context sensitivity of preferences, which is fairly intuitive (e.g., "in summer, I prefer ice cream over hot chocolate, and in winter, I prefer hot chocolate over ice cream). Additionally, it allows for multiple competing counterfactual beliefs about the effect of the law on behavior to combine to shape behavior. We committed to the context shaping view as the basis of our AIF model.

**2.2 AIF**

AIF is an agent-based modelling approach typically based on Partially Observable Markov Decision Processes (POMDP)[19,20]. POMDPs are used to model how sequences of decisions, actions or policies dictate the unfolding over time of states of a system and associated observations. For instance, a POMDP may be used to model robot navigation, by using as states the states of an environment (e.g., locations $l_1$ to $l_n$), and by using as outcomes the observations afforded by the location (e.g., observations $o_1$ to $o_n$). The POMDP can then be used to infer what policy should be selected to move the robot to the location that affords the preferred outcome, given a predetermined reward function, or outcome preference ranking. POMDPs can have multiple parameters but will minimally include parameters over the probability of transitioning between states given the possible reward maximizing actions the robot can take (a.k.a. transition probabilities), and parameters over the probability of observations in each state the robot can transition to (a.k.a. emission probability).

In AIF, POMDPs decompose into 4 basic sets of parameters, denoted as **A, B, C, D** and *G*. These relate the three fundamental variables of a POMDP: The outcomes or observations "o", which correspond to that which can be observed by the agent and are consequences of the states, the (unobserved, hidden or latent) states "s", which correspond to facts of the world, such as physical location and things generating observations, and "policies" denoted as "", which correspond to possible actions, or choices of the simulated agent that allow transitions between latent states.

The **A** parameter encodes the probabilistic relationship — the likelihood (P(o|s)) — that some outcomes will be found at certain locations, or states of the world. The policy dependent transition probabilities **B** (P($s_{t+1}$ | $s_t$, )) refer to the possible actions under each policy that can be engaged by the agent, and the **C** parameter encodes the — negative log probability of — "preferences" (-lnP(o)). Preferences in AIF are consistent with the ELA definition above, where **C** constitutes an order, or ranking R over some domain, where the R is over the outcomes, or observations or properties of states available in a given domain. The construct of "context dependent" preference is implemented in AIF as the conditionalization of preferences on context, where instead of being encoded as a matrix **C**{outcome modality}(outcome,timepoint) = [matrix]  or vector C{outcome modality}(1, outcome) = [vector], the preferences are encoded as a tensor **C**{outcome modality}(outcome, context), where the context corresponds to a latent state. The **D** parameter encodes the initial state probability of hidden states, and the *G* parameter — known as expected free energy — is used as the basis of a prior over policies (P()) and rests upon current beliefs about latent states and preferences.

In AIF POMDPs, the states "s" along with the policies "" have to be inferred. The observations are either observed or "predicted", or "generated" when inferring the action policies and the future observations expected under those policies. Additional variables can be inferred and updated. Here we will focus on only one such variable, the precision of beliefs about the most likely policy to pursue (gamma). This precision has been associated empirically with dopaminergic discharges in human neuroscience. Mathematically, it is the average expected free energy *G*. This policy precision scores the confidence of the agent over its policy[19]. When gamma attains its maximum of 0, this means that the agent trusts its assessment of policies and is certain about what to do next (for details see[19]).

Taken together, **A,B,C,D** and *G* allow for an inference of "where one should go— with a certain level of confidence (gamma) — and thereby what one should do in order to obtain preferred outcomes, given where one is starting from" and the current observation. This is achieved using several standard belief update equations in AIF that we briefly describe in Figure 1 (for a detailed description, see[19]).

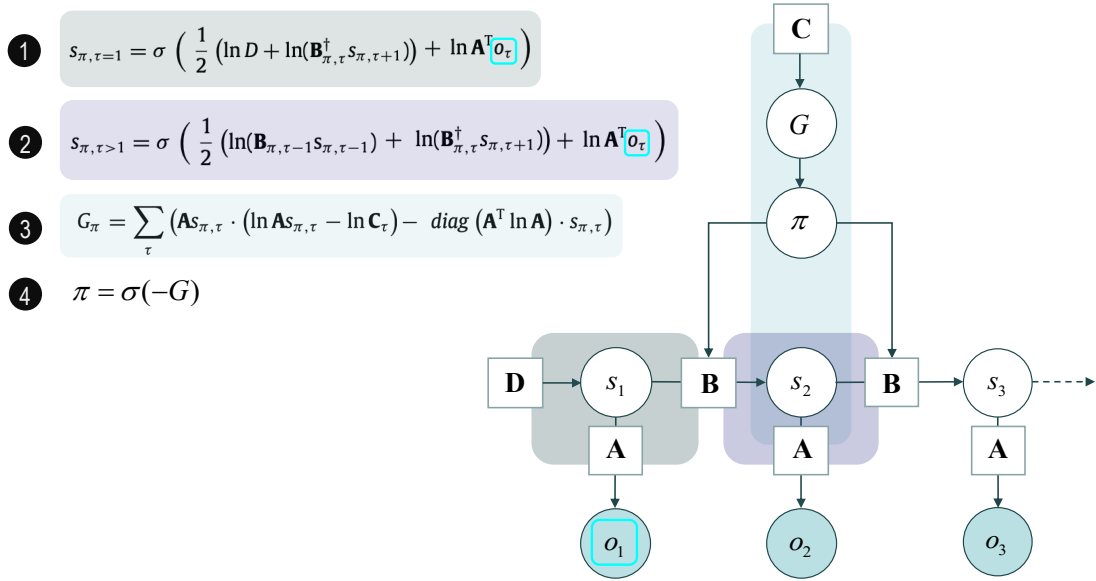

**Figure 1.** Inferences as they pertain to the **A,B,C,D**, and *G* parameters of POMDPs under AIF. This figure should be read in conjunction with(Smith, Friston, and Whyte 2022) for details, and noting that the equations in the figure refer to operations that can be implemented automatically in MATLAB using

routines in the SPM 12 toolbox. The color shaded areas grey, purple, and blue are there to help the reader identify the parameters involved in the equations. The dagger next to **B** and T next to **A** refer to transposition and (·) refers to a dot product. The greek letter tau refers to an index of the time — past, present, or future — where the present time always corresponds to the time of the observation currently under consideration. The term "diag" refers to the MATLAB function diag() that indexes and rearranges the diagonal elements of a matrix in a row vector. Square nodes refer to parameters, circle nodes refer to variables, with the filled circle node referring to observed or generated observation variables. Upon receipt of an observation $o\_t$, which is used to index a row in the likelihood matrix **A** (cyan square) one combines the initial states vector **D** with the transition matrix **B** and the indexed likelihood following the equation (1) to obtain a first estimate of the posterior over states, for each policy at the current time of observation. Plugging the **A** and **C** matrices into equation (2) allows for state inference over subsequent time points. This process is repeated iteratively, for each policy-dependent matrix, until the posterior distributions over states for all time points converge. Based on this inference about the past, present and future, the agent can then compute what the best course of action is, according to equation (3). Equation three produces a vector of expected free energy values $G$ for all possible policies, under prior preferences **C**. Expected free energy — as described in (3) — can be rearranged in several ways, each describing imperatives that an agent conforms to by virtue of selecting expected free energy minimal policies. Equation (3) reflects a decomposition in terms of "risk" and "ambiguity", where the risk corresponds to the divergence between the preferred outcome distribution and the predictive posterior over outcomes in the future, and where the ambiguity component refers to the conditional entropy of possible observations; given hidden states. By selecting policies that have low expected free energy, agents thus end up acting in a way that minimizes risk: i.e., brings them closer to outcomes that rank higher in their preferences, while minimizing the ambiguity of these outcomes. In the context of the model that interests us, the prior beliefs over policies () are calculated as the softmax (sigma; i.e., normalized exponential) function of the expected free energies $G$, as described in equation (4).

**2.3 Context shaping and the construct of context dependent preferences in AIF**

In AIF models, the observations and states should replicate — with as much fidelity as possible — the structure of the world, or generative process causing observations for the action to effectively be preference maximizing. For that reason, the representation of observations and states can be enriched by modelling multiple kinds of states and observations called state "factors" F and observations "modalities" M. State factors allow for the representation of world states with multiple causes: e.g., the color red might be caused by several different objects (F1) at different locations (F2). Complex causes can also generate observations in multiple observation modalities, where the modality can correspond to objects being seen or properties of those objects that can be sensed (e.g., via sight (M1), smell (M2), touch (M3), taste (M4), and hearing (M5), etc.). Factors and modalities relate through the likelihood parameters. When there is more than one factor, the likelihood is modelled as a tensor encoding the probability of observation in one modality (e.g., M1) conditional on hidden states for all the factors (e.g., F1 and F2) (P(M1 | F1, F2, …Fn). When considering multiple modalities, the model includes multiple likelihood tensors, one for each modality.

This way of modelling the structure of the world is interesting for our purpose because it captures the effect of observable and unobservable causes on action selection. For instance, a state like a location can be both a "hidden" cause to be inferred and an "observable" (i.e., "I infer where I will go" and "I can see where I am"). However, something like a "legal rule" is an unobservable cause of observations; especially those observations generated by other law-abiding agents. Hidden or latent states shape observations implicitly, by shaping the "context" within which observations become possible, and therefore become preferred to a greater or lesser degree. A latent context can be social, cultural, etc., but also legal. For instance, one may be less likely to observe oneself crossing an intersection when the light is red than when the light is green; and indeed, may be averse to making such an observation. This is so because penal codes contextualize our behaviors, preferences, and the very possibility of observations in the world without us having to carry a pocket version of the code at all times.

AIF thus offers a unique way to model latent legal causes in a way that reflects the "context shaping" view discussed in the previous section. Accordingly, we propose that a "legal context" in AIF can be modelled as a factor that contextualizes the preferences over observation modalities. For instance, upon observing a "full" centerline (cue M1), the preferred observable "location" for a driver (location M2) may be its current location (location F1) due the legal context established by a traffic code (context F2). In this case, the preference maximizing action is to "stay" (i.e., transition from current lane location to the current lane location). This is the kind of lawful behavior and preference one should be endowed with in a situation where the legal context mandates to "keep your lane" (legal context F2).

However, sometimes, the lawful conduct may be to override one rule in order to comply with another one, which may be more abstract (e.g., "yielding the right of way" by driving across a full traffic line to let an emergency vehicle pass). More fine-grained, and therefore human-like legal decision making depends on multiple layers of context prescribed by the law — and or cultural and social norms – shaping our preference for outcomes to be realized through action, and are induced through what has been called in the literature on AIF "deontic cues"[21]. Deontic cues trigger normed behavior by allowing one to zero in on contextually appropriate preference sets, through adjudicating between several competing layers of norms — legal or else — forming a counterfactual "if then" structure. Layers of conflicting legal contexts can be added to a POMDP generative model as additional state factors (F2, F3, …), thereby allowing for nuanced legal decision making akin to what can be observed in humans. We present a numerical (simulation) study in the next section to illustrate such an AIF approach to contextual legal decision making.

**3 SIMULATION STUDY**

**3.1 The task**

The purpose of our simulation is to show how AIF allows for intuitive human-like responses in a situation where there is a conflict of norms and wherein more sophisticated legal decision making should apply. In our simulation scenario, an autonomous vehicle operated by an AIF driven system has to decide whether it should cross from the left to the right lane, given 2 layers of normative contexts at hand. Both contexts are legal. The first context determines the permissibility of changing lanes (i.e., stay or cross) such as deontically cued by the nature of the centerline (i.e., full or dashed). The second context determines the permissible conduct in an emergency situation, deontically cued by a siren, and which may involve acting in contradiction with first order norms. In our simulation such a decision is implemented as "yielding the right of way" to an emergency vehicle (or complying with the "move over law"). Importantly, complying to second order norms, in this case, involves a trade-off; that between complying to the second order norm at the cost of violating a first order norm, which may trouble other denizens and lead to being honked at. This reflects, in a loose sense, the challenge that underwrites commonly modelled situations in the domain of self-driving cars (e.g., "crosswalk chicken"[22]).

We show that normatively appropriate behavior emerges from the fact of being endowed with context dependent preferences. We illustrate how one set of context dependent preferences can be overwritten by potentially conflicting preferences, in terms of choice behavior. In our simulation, the agent moves from a starting location (e.g., current position in lane in a start location 1) to a decision point position in location 2, and then to a location enacting the decision (e.g., to cross over the right lane, which is location 3). The agent can either stay and keep its lane in location 2, or after crossing through location 3 end up in a target location 4, on the right lane. Location 2 can be understood as a "in dilemma" zone[23]. In normal circumstances, the agent will cross the road when the centerline is dashed and will keep its lane when the line is full. However, in an emergency situation such as cued by a siren, a context dependent preference shift occurs, which can trigger the imperative to "yield the right of way" by moving on the right-hand side. This, however, exposes the driver to an aversive outcome (e.g., honking from other drivers) if the line is full, and therefore can only occur if the situation is indeed an emergency one.

In our simulation, agents make 10 sequential decisions and can plan over 4 time steps into the future, and therefore can plan over the entire sequence of 4 possible states. We run our simulations under 7 conditions controlling for the two layers of normative context by supplying the context states — over which the driver has no control — to the model, by adding a MDP.s structure to the MDP to be passed through the function SPM_MDP_VB_XXX. The 7 scenarios we simulated are as follows (see box 1 for the code):

*Full line (stay in lane) conditions*
C1: Stay in lane over the 10 decisions and no emergency over the 10 decisions;
C2: Stay in lane over the 10 decisions with an emergency over the 10 decisions;

*Dashed line (cross lane) conditions*
C3: Cross the line over the 10 decisions and no emergency over the 10 decisions;
C4: Cross the line over the 10 decisions with an emergency over the 10 decisions;

*Mixed lines (stay/cross) conditions*
C5: Mixed stay/cross over the 10 decisions and no emergency over the 10 decisions;
C6: Mixed stay/cross over the 10 decisions with emergency over the 10 decisions;
C7: Mixed stay/cross over the 10 decisions with mixed over the 10 decisions.

**Box 1.** Code to supply in order to simulate the 7 conditions using DEM.law as per the software specification note

```
%%% Condition 1:Full line X no emergency
MDP.s     = zeros(Nf,MDP.T);      % Location (chosen lane)
MDP.s(2,:) = [1 1 1 1 1 1 1 1 1];  % Context 1 (stay/cross)
MDP.s(3,:) = [1 1 1 1 1 1 1 1 1];  % Context 2 (normal/emergency)
%%% Condition 2: Full line X emergency
MDP.s     = zeros(Nf,MDP.T);      % Location (chosen lane)
MDP.s(2,:) = [1 1 1 1 1 1 1 1 1];  % Context 1 (stay/cross)
MDP.s(3,:) = [2 2 2 2 2 2 2 2 2];  % Context 2 (normal/emergency)

%%% Condition 3: Dashed line X no emergency
MDP.s     = zeros(Nf,MDP.T);      % Location (chosen lane)
MDP.s(2,:) = [2 2 2 2 2 2 2 2 2];  % Context 1 (stay/cross)
MDP.s(3,:) = [1 1 1 1 1 1 1 1 1];  % Context 2 (normal/emergency)

%%% Condition 4: Dashed line X emergency
MDP.s     = zeros(Nf,MDP.T);      % Location (chosen lane)
MDP.s(2,:) = [2 2 2 2 2 2 2 2 2];  % Context 1 (stay/cross)
MDP.s(3,:) = [2 2 2 2 2 2 2 2 2];  % Context 2 (normal/emergency)

%%% Condition 5: Mixed line X no emergency
MDP.s     = zeros(Nf,MDP.T);      % Location (chosen lane)
MDP.s(2,:) = [1 1 1 1 1 1 2 2 2];  % Context 1 (stay/cross)
MDP.s(3,:) = [1 1 1 1 1 1 1 1 1];  % Context 2 (normal/emergency)

%%% Condition 6: Mixed line X emergency
MDP.s     = zeros(Nf,MDP.T);      % Location (chosen lane)
MDP.s(2,:) = [1 1 1 1 1 1 2 2 2];  % Context 1 (stay/cross)
MDP.s(3,:) = [2 2 2 2 2 2 2 2 2];  % Context 2 (normal/emergency)

%%% Condition 7: Mixed line X Mixed emergency
MDP.s     = zeros(Nf,MDP.T);      % Location (chosen lane)
MDP.s(2,:) = [1 1 1 1 1 1 2 2 2];  % Context 1 (stay/cross)
MDP.s(3,:) = [1 1 1 2 2 2 2 2 2];  % Context 2 (normal/emergency)
```

### 3.2 The generative model

The generative model includes 3 state factors (F1,F2,F3) and 4 observation modalities (M1,M2,M3,M4). See Figure 2 for a summary. The factors are: (F1) location or lane, (F2) normative context 1 (i.e., permissible conduct under the law, such as driving within lane or crossing), and (F3) normative context 2 (i.e., permissible conduct under the law in a state of emergency, such as yielding the right of way). F1 states are locations 1 to 4, F2 states are "stay" (i.e., keep lane) and "cross" (i.e., yield by crossing), and F3 states are "normal" and "emergency". The modalities are: (M1) observed locations, (M2) deontic cue 1; (M3) deontic cue 2, and (M4) deontic cue 3 (i.e., fellow drivers' signaling, such as honking on/off). M1 includes the observations of locations 1 to 4, M2 observations are the "full" or "dashed" line, M3 observations are the state "off" and "on" of the siren signaling the presence of an emergency vehicle, and M4 observations "off" and "on" sounds generated by fellow drivers.

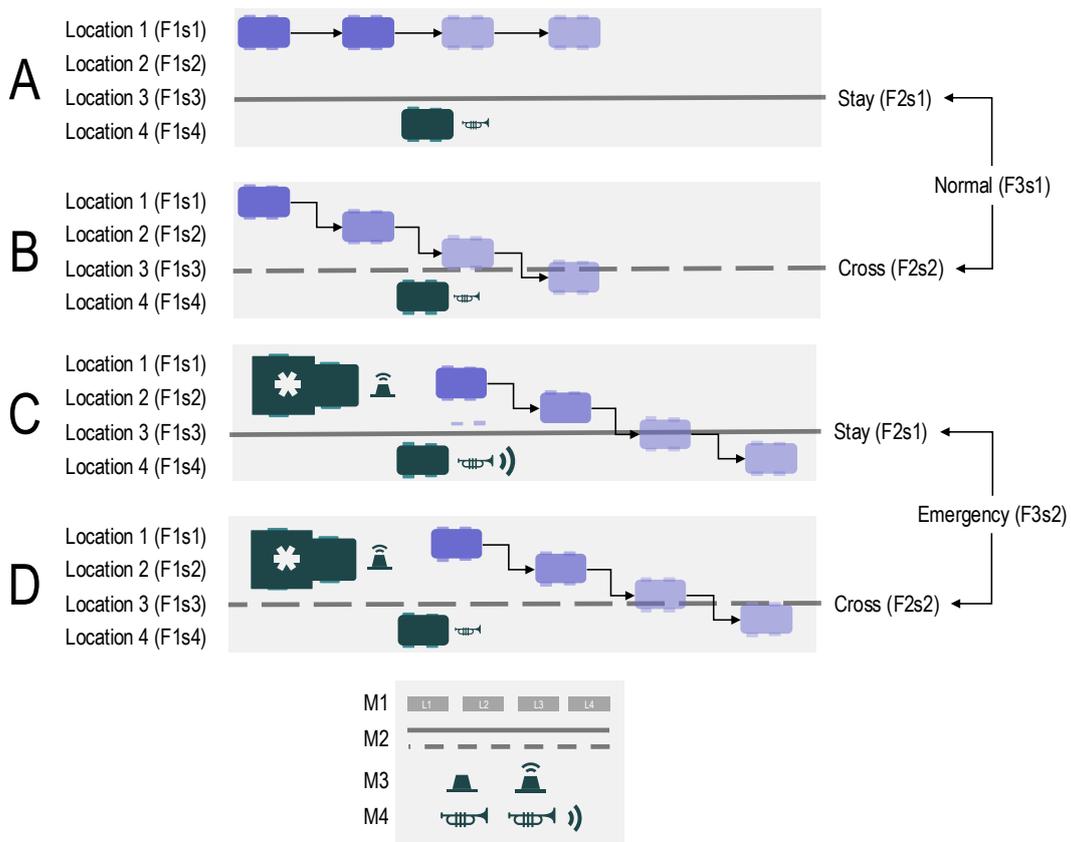

**Figure 2.** Depiction of the factors and modalities. Bottom: Depiction of the 4 modalities. **Panel A**: Depiction of the behavior expected in normal context 1, for the "stay" context 2 (i.e., keeping one's lane until the dash line appears). **Panel B**: Depiction of the behavior expected in normal context 1, for the "cross" context 2 (i.e., changing lane). **Panel C**: Depiction of the behavior expected in emergency context 1, for the "stay" context 2 (i.e., changing lane). This behavior violates the norm under context 1, and is enabled by preference context dependency. **Panel D**: Depiction of the behavior expected in emergency context 1, for the "cross" context 2 (i.e., changing lane). This complies to both normative contexts.

The specific parameterization of the generative model is detailed in figure 3. The cue 1 (i.e., traffic cue) modality (A,1, fig 2) includes fully uncertain mappings under all contexts at locations 1 (i.e., start) and 4 (i.e., target), and fully certain mappings under all contexts at locations 2 and 3. These precise mappings are inverted depending on whether the agent finds itself in the "cross" or "go" state of context 1. The likelihood of "full" in "stay" state is 100%, whereas the likelihood of "dashed" in "cross" state is 100%. This ensures that the agent sees "full" when the rule requires a "stay" and "dashed" when the rule allows for a "cross", irrespective of context 2.

The cue 2 (i.e., siren modality (A,1, fig 3) includes high certainty mappings (87.5%) for the "off" observations under both states of context 1 (stay/cross) and under the "normal" state of context 2. This mapping is then inverted under the "urgent" state. This means that the agent will most probably not hear a siren in the normal case and will most probably hear one in the emergency case. The cue 3 (i.e., alert) modality (A,1, fig 4) defines identical mappings under each possible states of context 2 (normal/emergency), but that is different depending on whether the agent is in a "stay" or "cross" state in context 1. In the "stay" context, the agent does not hear honking at locations 1, 2 and 4, but hears one at location 3, and in the "cross" context simply never hears it.

Narratively, the belief structure of the agent — as defined by the generative model — is that the agent:

(i) always sees where it is at (i.e., "I always know where I am");

(ii) always perceives the full or dashed line in a way that is coherent with the legal context 1, when at locations 2 and 3, and remains uncertain at location 1 and 4 (i.e., "I always see the line appropriate to the first order legal context, when this line becomes visible to me, that is when I am about to cross and while crossing");

(iii) most probably hears the alarm when contextually appropriate, irrespective of whether it should "stay" or "cross" (i.e., "I will most probably hear the siren in the urgent context, and most probably not in normal context");

(iv) is never honked at, except under the crossing state (i.e., location 3) and in the "stay" state of context 1 (i.e., "the only time I hear honking is when crossing when I should have stayed in lane").

In short, the agent believes that "I see what lane I am in", "I see that I should stay or cross when the line is full or dashed", "I tend to hear sirens when there is an emergency", and "I hear people honking at me when I override elementary traffic rules".

The transition probabilities over states in F1 are specified for two policies (Figure 3, B, 1). Policy 1 allows for transitioning from each state to itself, thereby implementing the "stay" policy. Policy 2 allows for transitioning from state 1 to 2, from state 2 to 3, from state 3 to 4, and from state 4 to 3), thereby specifying a "steer" policy. The transition probabilities over states in F2 and F3 are specified for one policy each that allows for respective context switching (fig 3, B, 2 and 3). The initial state vectors **D** for each factor are represented in Fig. 3, **D**. The vector for F1 indicates the agent believes with 100% probability that it starts at state 1 (Fig. 3, **D**, 1), that the legal context remains uncertainty (Fig. 3, **D**, 2), and that the legal context 2 is mildly uncertain, with 87.5% probability attributed to the "normal" state context, and 12.5% probability attributed to the "emergency" state context.

When parameterizing an AIF POMDP, one can specify preferences over the outcome space. The specific parameterization of our simulation is detailed in Figure 3, **C**. The structure of the preference is the same as that of the modalities. What changes is the parameterization. Following the preference set for the first modality (Fig. 3,C,1), the agent has a mild preference for the 4th, target location when in normal state context, and a much stronger preference for the target state when in the urgent state context. This means that the agent will experience a preference shift when inferring that it is in an urgent context, upon hearing a siren sound. The only other nonzero preferences are those for the alert cue modality (e.g., honking by other drivers). Irrespective of the contexts, the agent has a significantly higher preference for not hearing an alert sound. This effectively plays against decisions of "yielding the right way" when in a legal context where the agent should keep its lane, such as evidenced by the full line deontic cue. The only way this aversion to honking can be overcome is in the emergency context.

Note that this specific parameterization has been selected for the purpose of illustrating the potential of AIF as an agent-based modelling approach to normative behavior. Parameters of an AIF agent can be learned[24] based on observed outcome frequencies, which could allow for the automatic acquisition of a parameterization reflecting specific local cultural norms.

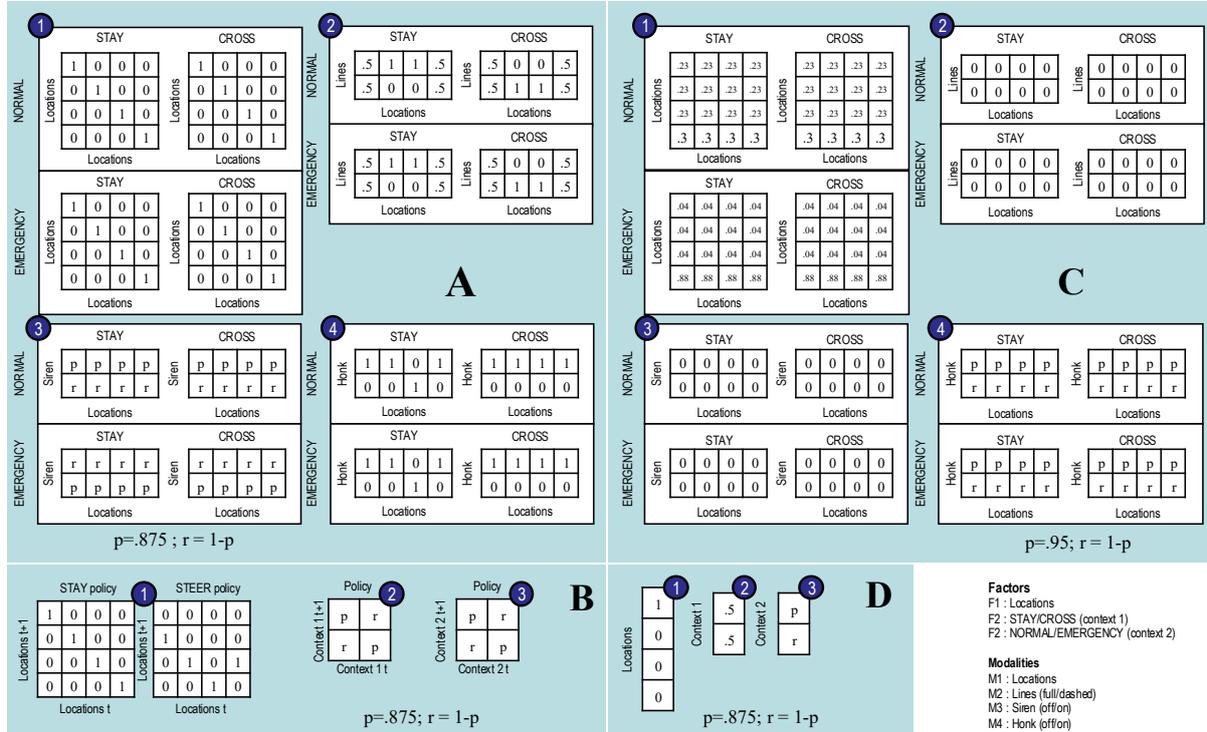

**Figure 3.** Beliefs and observation spaces for the POMDP. Panel **A** presents the likelihood tensors, panel **B** presents the transition tensors and matrices, panel **C** presents the context dependent preference tensors that allow, and panel **D** presents the initial state vectors.

### 3.3 Results

The choice behavior under the 7 conditions are summarized in Figures 4 to 7, with Figure 4 presenting results of conditions 1 and 2, Figure 5 presenting results of conditions 3 and 4 , Figure 6 presenting results of conditions 5, 6, and 7. We discuss the results each in turn. The figures are separated into 5 panels, along 3 rows and 2 columns. These panels describe:

**(i) Panel 1,1:** the inferred hidden states (red dot) for each state factor (locations, legal context 1, and legal context 2) over the 10 time steps in the foreground and the beliefs in the background (black = 1, white = 0, grey = range >0<1);

**(ii) Panel 1,2:** the inferred action policy (cyan dot) (i.e., stay, or steer) at each time step in the foreground and beliefs about locations in the background;

**(iii) Panel 2,2:** the posterior beliefs over policies throughout the belief updates process;

**(iv) Panel 3,1:** the generated outcomes (cyan dots) laid over the outcome preference in the background, for each modality;

**(v) Panel 3,2:** the confidence or precision over policies (i.e., the negative average of expected free energy *G*). The cyan line plots the value of gamma, and the black bars chart the rate of change of gamma. When modelling neurobiological processes with AIF, these are used to simulate neurophysiological responses (e.g., dopaminergic response[25]). In the context of an autonomous vehicle, gamma does not have a neurophysiological correlate. It should simply be viewed as tracking the AI agent's confidence over its selected policy, and indirectly the level of "vigilance" of the autonomous vehicle.

*Conditions 1 and 2*

In condition 1, the agent only moves one time to get into the "dilemma zone", and stay in that zone for the remainder of the trial. This is due to the fact that the legal context 1 cued by the full line requires that the agent avoid changing lanes, and that the legal context 2 is "normal". No siren is heard, and nobody honks at the agent; hence, uncertainty remains over the posterior probability of policies, such as reflected by the expected precision. This is due to the fact that the agent maintains some degree of uncertainty over the context it is in, which otherwise would be disambiguated by the sound of a siren indicating an emergency. While performing the normatively appropriate behavior, the agent maintains low confidence over its choice. The lack of confidence over policies evinces a feature of cue sensitive agents: in the opposite scenario, where say, the agent would be context insensitive (i.e., would have full certainty over one of the 2 contexts), such an agent would not be able to make the kind of nuanced, context sensitive decision that we expect from human normative agents. A context insensitive agent would only act according to one context only, either always thinking that breaking basic rules is permitted because of a constant context of emergency, or being too rigid to adjust on the fly upon hearing an alarm sign. Low confidence is thus conducive to the kind of vigilance that allows for normatively appropriate conduct in context.

In condition 2, the agent moves three times from the beginning, to move from the start location to the "dilemma zone" and then to the crossing state and to the target state. This is due to the fact that despite the legal context 1 requiring that the agent keep its lane, the "emergency" in legal context 2 licenses the agent to cross in order to "yield the right of way". The agent crosses despite getting honked at, confidently, knowing that it is licensed to do so in a state of emergency, as shown by the posterior probability over policies and expected precision that spikes at the time of crossing the lane.

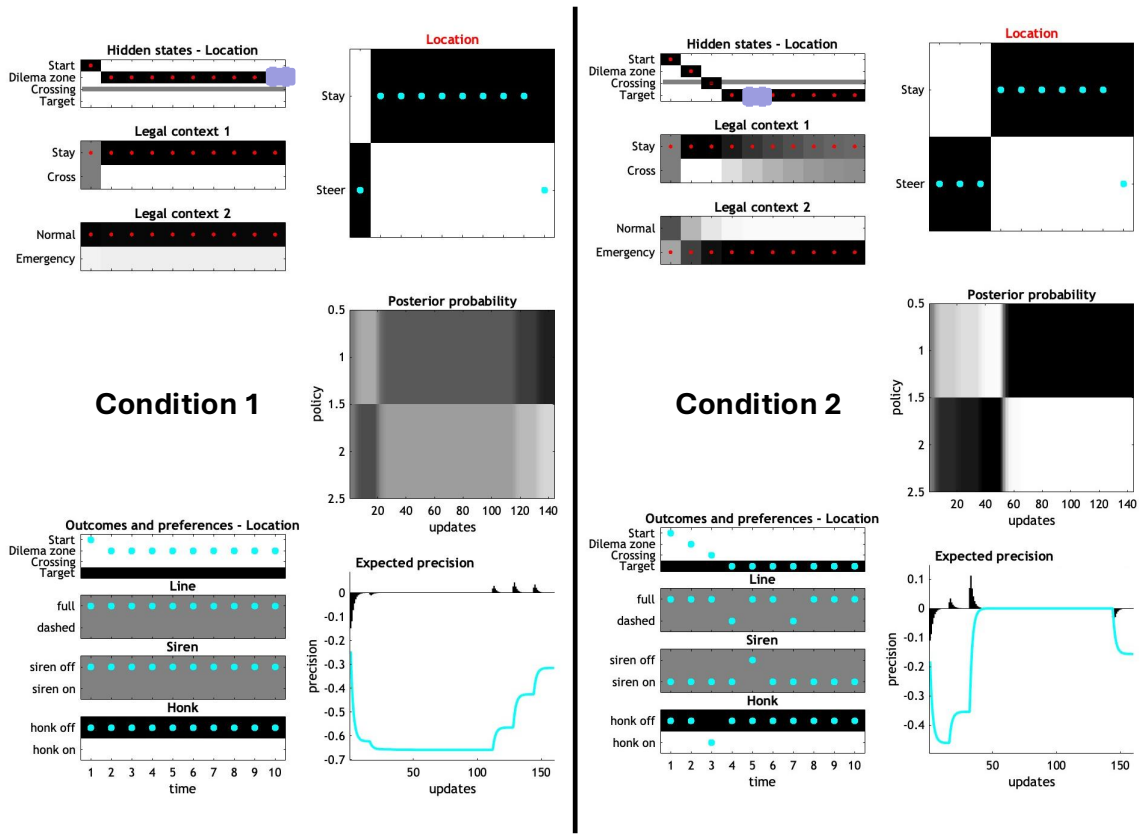

**Figure 4.** Results for conditions 1 and 2.

*Conditions 3 and 4*

In condition 3, the agent also moves three times from the beginning to move from the start lane to the target lane. This is fully licensed under legal context 1 (cross), such as cued by the dashed line. Note however that the agent is not confident about its decision, because it remains uncertain about the context it is in. In condition 4, the agent displays the same behavior as in condition 3, but with progressively increasing confidence, again, induced by the siren cue.

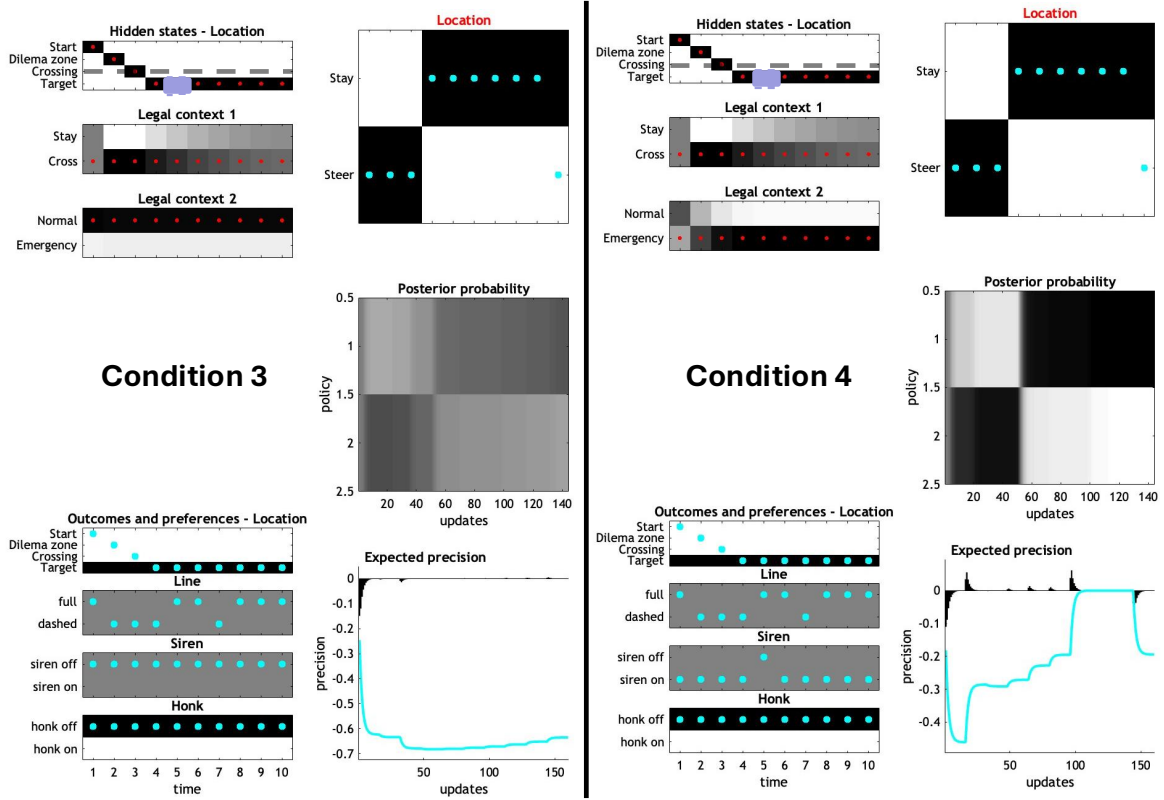

**Figure 5.** Results for conditions 3 and 4.

*Conditions 5, 6 and 7*

In condition 5, the agent first steers into the dilemma zone to then cross the centerline when it becomes dashed. This is presented with a normal context, which licenses it to cross only over a dashed line. Confidence remains low throughout, meaning that the agent remains vigilant and poised to act appropriately in an emergency. In condition 6, the agent quickly crosses over to the other side upon hearing the siren, despite the line being full; thereby taking the risk of, and indeed being, honked at. Similar behavior emerges under condition 7, with the agent remaining vigilant slightly longer due to the delayed disambiguation of the emergency upon crossing the centerline.

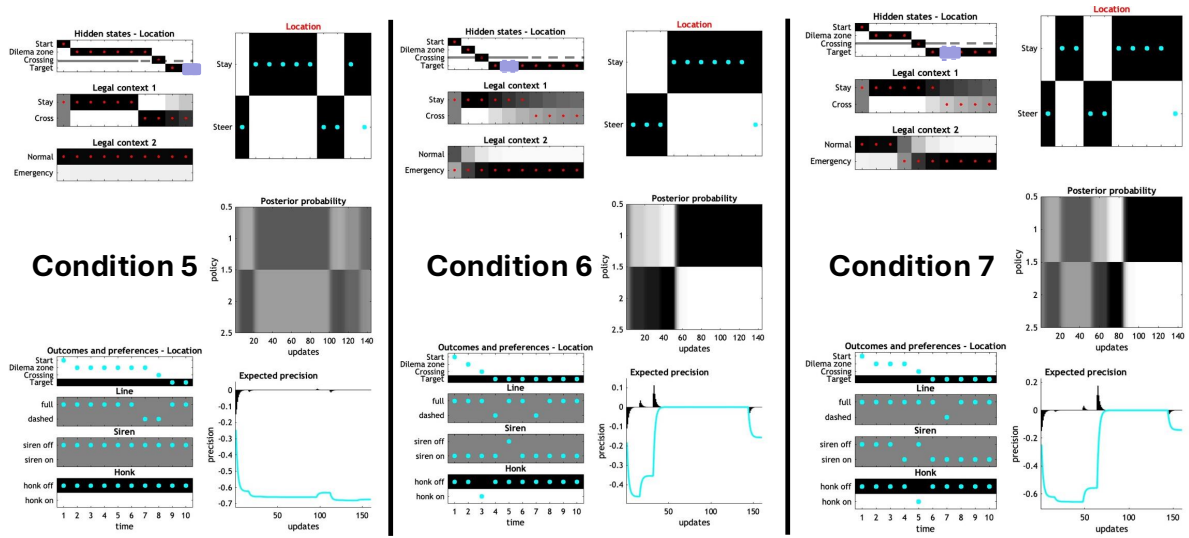

**Figure 6.** Results for condition 5 and 6

### 3.4 Discussion

The goal of this paper was to show how context dependent preferences allow AIF driven agents to act adequately in context, when facing a normative conflict, in ways that resemble what is expected from human agents (e.g., what is expected according to traffic rules and the obligation to "yield the right of way"). We have tried to establish the construct validity of context dependent preferences by grounding them in a rationale derived from ELA theory. Additionally, our numerical study speaks to the face validity of the construct. One limitation of our approach is that our model has to be designed to accomplish a specific task and is not designed to learn from experience. Our model should be viewed as providing a bespoke "normative module" that an AI agent could leverage in specific situations requiring more sophisticated normative decision making (e.g., in a "yielding the right of way" situation).

Although task specificity limits the flexibility of our model, parameterization can be learned thereby providing additional degrees of flexibility to match local cultural, task-specific normative expectations (e.g., by adjusting the preferences over locations involving crossing over a full line based on observed frequencies of human conduct). For instance, in a pedestrian simulation scenario, this might mean adjusting the preferences for crossing, or not, on a red traffic light when there is no one around, based on what is observed culturally (e.g., in France people tend not to bother with traffic lights if there is no risk, whereas in Germany people may tend to respect signalization rigidly). Additionally, the construct of context dependent preference is in itself sufficiently general to accommodate any normative conflict for any kind of norms requiring that, fundamentally, a behavior P be permitted or not relative to one or more normative contexts Q at hand, upon receipt of an observation input indicating the context. Although not explored in the above numerical studies, this kind of person-specific preferences is implemented in a straightforward way by adjusting the precision of prior preferences in **C**. This affords the opportunity to introduce not only context sensitivity in terms of reversing preferences but also in terms of reversing the rank, where precise preferences predominate over less precise preferences. A complementary application arises in the context of computational phenotyping; namely, finding the precisions of preferences that best explain somebody's choice behavior. This is an established procedure in computational psychiatry, where the aim is to quantify and characterize patients (or cohorts) in terms of their prior beliefs about the way they should behave[26,27]

Our simulations suggest that an agent's precision "gamma" (i.e., precision of the distribution over allowable policies) changes as a function of deontic cue disambiguation. Two generic patterns emerge: (i) low precision (i.e., confidence) under unresolved or ambiguous normative context (conveying heightened vigilance), and (ii) transient gamma spikes when context is disambiguated (e.g., a siren licenses crossing a solid line) or when decisive action is selected in conflict contexts (felt "conviction"/relief). In our plots, Panel 3,2 explicitly tracks gamma (cyan curve) and its rate of change (black bars), and these traces covaried with policy posteriors during legal dilemmas.

Interestingly, the dynamics of gamma have been argued to reflect the affective aspect of belief updating in human subjects, where valence and arousal emerge from precision-weighted prediction-error flow and belief updating about policies[28,29]. In our simulation, low gamma under unresolved legal context (e.g., solid line, no siren) corresponded to high arousal / cautious vigilance and negative/uncertain valence. Gamma rose as deontic cues (siren) resolve conflict, corresponding phenomenologically to relief or conviction once the agent infers that crossing (even with social sanction, i.e., honking) is licensed, and gamma trajectories and posterior mass shifts at the moment of crossing in emergency conditions (Condition 2; compared to Condition 1), and in Conditions 4, 6–7.

In AI agents, it is unclear that gamma dynamics can be said to track something like affect. However, it does function as an indicator that one can use to quantify the confidence an agent has over its action space; just like one would do with affect in the human case. In the context of driving, for instance, affective cueing is crucial when interacting with other drivers, where the affective response of other drivers – such as by facial expression and bodily attitude — can be used to disambiguate a situation (e.g., having to decide which of multiple cars arriving at the same time at an intersection should yield). Policy precision, such as modelled here, could be displayed, say, in the form of a colored light on top of a self-driving car to indicate the "affective" state of the car and thereby help human (or other AI) drivers make better decisions: e.g., deciding to letting the right of way at the intersection upon noticing that the other car has high confidence over its policy space, in a context where it is conceivable that the car would decide to accelerate.

## 4. CONCLUSION: AIF FOR AI GOVERNANCE?

In the introduction, we implied that our model could function as a "safety valve" of sorts for regulation by design approaches to AI governance: the actor and the agent approaches. Having detailed the working of our model — and justified its operations in the light of ELA – we conclude by returning to issues of AI governance and how "normative modules" can help in mitigating risks posed by AI agents.

Governance, broadly construed, refers to the steering of the behavior of an individual, a group of individuals, or a set of state institutions (e.g., members of a society, or organs of the state)[30] towards the delivery of goods (e.g., public goods)[31]. Accordingly, AI governance can be read as the steering of artificial intelligence software towards its delivery as a good, where the delivery can be considered appropriate, if it is done by mitigating the different risks posed by AI systems and by its industry.

Actor governance corresponds to what is sometimes called "organizational" governance[32]. Actor governance concerns the way the actors of the AI value chain, from suppliers of the hardware to deployers of models, govern their activity towards the delivery of AI software while mitigating the risks posed by the delivery of those software. These risks include socio-economic and geopolitical risks caused by imbalances in the AI value chain (e.g., the outsourcing of low value work like labelling to countries with less control of the value chain), the technical risks of AI that arise along the product life cycle (e.g., transparency, explainability, fairness, robustness, etc.), and the environmental risks of AI associated with impacts of the operational costs of AI production[33].

In turn, agent governance concerns mitigation of risks posed by the conduct of autonomous AI systems such as robots, drones, or Internet of Things (IoT) devices[34,35]. The AIF normative module developed in this paper is relevant primarily for agent governance purposes. Agent governance is about ensuring that decisions of autonomous or semi-autonomous AI systems remain aligned with human defined norms, legal or ethical. Strategies of agent governance function as proxies to the minimization of AI misalignment risks[36] posed by the behavior of autonomous AI systems. Misalignment risk include[37]:

**(i) Planning risks:** risks associated with the capacity of autonomous AI systems to plan decision over long time horizons, which may lead, for instance, to the manipulation of users' beliefs;

**(ii) Empowerment risks:** risks related to empowerment of AI agents by humans or humans' overreliance on autonomous AI systems (e.g., when considering outsourcing policy making to AI),

**(iii) Unidentified risks:** risks that stem from unpredictable harm caused by emergent behavior.

Misalignment risks are thought to stem from 4 agentic capacities[37]:

**(i) Underspecification:** the capacity to accomplish a goal in the absence of human specification of how the goal can be accomplished;

**(ii) Impact:** the capacity to affect the environment without human intervention;

**(iii) Goal-directedness**: the capacity to achieve objectives seemingly autonomously;

**(iv) Planning:** the capacity to make coordinated decisions over long time horizons.

Misalignment risks, of course, can be mitigated with actor governance strategies, such as the adoption of design principles by AI systems providers guaranteeing interpretability (i.e., explainability of the decision making process), controllability (i.e., possibility of keeping a "human in the loop"), and ethicality (i.e., the compliance to humans defined norms)[38]. However, built in mechanisms for behavioral control should also be employed to mitigate agent governance risks at their source. One can imagine how an AIF normative module could be used to mitigate misalignment risks through enabling the agent to "self" constrain its capacities based on applicable norms. For instance, such a module could be used to:

**(i) self-constrained under-specification:** defining contexts under which different conducts are permissible. This does not limit the agent in its ability to find a solution to a problem on its own. It simply establishes boundaries on the solutions that can be found and ensures that all the solutions will align with human defined normative expectations (e.g., ensuring that deciding autonomously when the best moment to cross over the centerline is always within the logic of applicable norms);

**(ii) self-constrained impact:** ensuring that the impact of an agent's actions is mediated by norms. This does not reduce the impact, but dampens it by aligning the impact with human normative expectations (e.g., an autonomous car may decide to yield the right of way and in so doing cause a crash, but this crash would be expected and explainable at the light of what a legally rational human would have done);

**(iii) self-constrained goal-directedness:** allowing for preference maximizing behavior to be context sensitive thereby allowing for change in preference as a function of the normative demands of the situation (e.g., shifting from having a mild preference for the target lane location to a stronger preference for the target lane location in an emergency context). This is not about limiting the preferences of the agent. It is about contextualizing them to normative expectations.

**(iv) self-constrained planning** by factoring into plans information about the normative context, to ensure that all plans align with normative expectations (e.g., ensuring that all policies involve actions that negotiate applicable norms, irrespective of the time horizon of the policy). This does not involve limiting long term planning capabilities, but rather adding "guardrails" around planning.

Mitigating agent governance risks, vicariously, through using an AIF normative module to align agentic capacities with normative expectations is, in our view, a promising approach. On the one hand, the "affective" implications of our approach means that one can design agents whose gamma updates are sensitive to norm cues, and thus obtain principled caution in ambiguous contexts and rapid, confident commitment when higher-order norms apply (e.g., emergency "move-over" rules). Our results explicitly show that low confidence is conducive to the kind of vigilance that allows for normatively appropriate conduct in context, which can be communicated to other agents, artificial or humans, to ensure harmonious interactions. On the other hand, our approach does not require limiting the capacities of AI agents (or slowing down the technical innovation of AI actors developing those agents), but simply requires implementing normative "guardrails" on AI conduct.